%% file: Pedrini_D0mumu_Charm2012.tex
\documentclass[12pt]{article}
\usepackage{graphicx}

\def\MIB{INFN Sezione di Milano-Bicocca, I-20126 Milano, ITALY}

\def\Title#1{\begin{center} {\Large #1 } \end{center}}
\def\Author#1{\begin{center}{ \sc #1} \end{center}}
\def\Address#1{\begin{center}{ \it #1} \end{center}}

\newenvironment{Abstract}{\begin{quotation}  }{\end{quotation}}
\newenvironment{Presented}{\begin{quotation} \begin{center} 
             PRESENTED AT\end{center}\bigskip 
      \begin{center}\begin{large}}{\end{large}\end{center} \end{quotation}}

\input econfmacros.tex
\begin{document}
\begin{titlepage}
%\pubblock

\vfill
\Title{Search for the flavour-changing neutral current decay $D^0 \to \mu^+\mu^-$ 
in pp collisions at $\sqrt{s} =$ 7 TeV with CMS}
\vfill
\Author{ Daniele Pedrini\\ 
on behalf of the CMS Collaboration}
\Address{\MIB}
\vfill
\begin{Abstract}
A search for the flavour-changing neutral current decay $D^0 \to \mu^+\mu^-$  is performed in pp collisions at 
$\sqrt{s}=$ 7 TeV using 90 pb$^{-1}$ of data collected by the CMS experiment at the LHC. No evidence is found for this decay mode. 
The upper limit on the branching fraction $\mathcal{B}(D^0 \to \mu^+\mu^-)$ is $5.4 \times 10^{-7}$ at the $90\%$ confidence level.
\end{Abstract}
\vfill
\begin{Presented}
Charm 2012\\ 
The $5^{th}$ International Workshop on Charm Physics\\
Honolulu, Hawaii, May 14--17, 2012
\end{Presented}
\vfill
\end{titlepage}
\def\thefootnote{\fnsymbol{footnote}}
\setcounter{footnote}{0}

\section{Introduction}

One promising way to search for physics beyond the Standard Model (SM) is to search for decay modes that are extremely rare
or forbidden. The observation of these modes at rates exceeding the prediction of the SM could open a window onto New
Physics (NP). The Flavour-Changing Neutral Current (FCNC) decays are rare decays which proceed via an internal quark 
loop in the SM but are forbidden at the tree level. In the SM, the FCNC decay $D^0 \to \mu^+\mu^-$ is highly suppressed by the
Glashow-Iliopolus-Maiani (GIM) mechanism and by a factor of $m_\mu^2/m_\textrm{D}^2$ due to helicity. This decay proceed via 
a {\it W} box diagram, which also contributes to $\mathit{D}^0-\overline{\mathit{D}}{}^0$ mixing. Theoretical estimates of 
this branching ratio are approximately $10^{-18}$ from short range processes, increasing to $10^{-13}$ when long-distance processes 
are included~\cite{Burdnam}. However NP models can enhance these estimates by several orders of 
magnitude~\cite{Golowich}. This is why these decays are so attractive: any detection, given current
sensitivities, will be a clear sign of NP. \@ Large Hadron Collider (LHC) experiments have the possibility 
to detect these rare decay modes. Furthermore, as charm is an up-type quark, the search for FCNC in the 
charm sector is complementary to {\it B} and {\it K} decay searches.

In this note, a search for $D^0 \to \mu^+\mu^-$ is presented using a data sample of pp collisions 
at $\sqrt{s} =$ 7 TeV, corresponding to an integrated luminosity of $\sim$ 90$^{-1}$, collected by the 
Compact Muon Solenoid (CMS) experiment at the LHC.

\section{Analysis}

The strategy of this analysis is to measure the ratio of branching fractions, 
$\frac{D^{*+} \to D^0(\mu^-\mu^+)\pi^+}{D^{*+} \to D^0(K^-\mu^+\nu)\pi^+}$, in such a way that most of the 
systematic uncertainties cancel out (throughout this note charge conjugate state is implied).
The main challenge of this analysis is to reconstruct the normalization mode, which has a much smaller trigger
efficiency.
A feature of heavy flavour events is to have a secondary vertex separated from the primary vertex. 
For this reason the semileptonic and the dimuon analyses are 
essentially topological: they are based on a search for primary and secondary vertices in the event.
In the case when multiple primary vertices are reconstructed, the vertex with the highest sum of
$p_T^2$ of the tracks in the vertex is selected, where $p_T$ is the transverse momentum. This is the 
default CMS choice.  Events without a primary vertex or events in which the selected primary vertex has a 
$\chi^2$ probability less than 1\% are discarded.
A detailed description of the CMS detector can be found in Ref.~\cite{CMS}. The main subdetectors
used in this analysis to reconstruct the topological configuration of the event are the silicon 
tracker (composed of pixel and strip layers) and the muon stations. 
The trigger plays an essential role. The events selected for this analysis are those
with a muon having a $p_T$ greater than a certain threshold, which varies with running conditions. 
The events are selected with a two-level trigger system. The first level requires a good quality muon, while 
the high level trigger (HLT), using additional information from the silicon tracker, imposes a cut on the $p_T$ 
of the muon (for example {\it HLT\_Mu3} means an event with a muon having $p_T > 3$ GeV/{\it c}). 
As the luminosity of the LHC increased the triggers were prescaled.
During the 2010 data taking, the lowest $p_T$ unprescaled 
trigger was varied six times from 3 ({\it HLT\_Mu3}) to 15 GeV/{\it c} ({\it HLT\_Mu15}). In the 2011, the 
{\it HLT\_Mu15} trigger remained unprescaled for the first 54 pb$^{-1}$ of collected luminosity.
Later CMS trigger configurations are too inefficient to reconstruct the normalization decay mode 
$D^0 \to K^- \mu^+\nu$. The data sample considered is therefore divided into seven run periods: 
six for the 2010 data taking and one for 2011. 

The $D^0 \to \mu^+\mu^-$ analysis begins with two opposite sign muon candidates.
Muon candidates are required to be reconstructed both in the silicon tracker and in the muon stations. 
The global fit must have a $\chi^2$ per degree of freedom (DOF) less than $3$, to have a distance of closest approach to 
the primary vertex in the transverse plane less than $2$ mm, and to be within the pseudorapidity region 
$|\eta| < 2.1$, where $\eta = -\ln{[\tan(\theta/2)]}$ and $\theta$ is the polar angle with respect to the 
counterclockwise beam direction. In addition, one of the two muon candidates must match the muon which 
fires the trigger ({\it trigger muon}); this implies the following $p_T$ cuts for the 
different 2010 run periods: $p_T>$ 3, 5, 7, 9, 11 and 15 GeV/{\it c}. 
The $7^{\mathrm{th}}$ run period is the 2011 data taking with $p_T>15$ GeV/{\it c}. The other muon 
({\it second muon}) must satisfy  $p_T>3$ GeV/{\it c}. The two muon candidates must 
form a vertex with at least a 1\% $\chi^2$ probability ($\textrm{CL}_\textrm{sec} > 1\%$). If a good 
secondary vertex is found, the position of the primary vertex is recomputed, excluding the two muons 
from the vertex. A point-back cut is applied: the angle between the $D^0$ momentum vector and the 
line of flight of the $D^0$ (direction between the primary and secondary vertices) must satisfy 
$\cos{\alpha} > 0.99$. Finally the $D^0$ candidate is combined with a track ($p_T>0.6$ GeV/{\it c}), 
which is given the pion mass and must originate from the primary vertex, to form a $D^{*+}$ candidate. 
Combinations with $\Delta M = M(\mu^+\mu^-\pi^+) - M(\mu^+\mu^-)$ exceeding $180$ MeV are discarded; 
if more than one candidate is found with $\Delta M < 180$ MeV/$\textrm{c}^2$, the candidate with $\Delta M$ closest to
the nominal PDG~\cite{PDG2010} mass difference is chosen. 

The semileptonic decay mode reconstruction, developed by E691~\cite{E691}, is based on the decay chain 
$D^{*+} \to D^0(K^-\mu^+\nu)\pi^+$ whereby the $\nu$ momentum can be reconstructed provided the $D^0$ direction 
(vector between the primary and secondary vertices) is sufficiently precise.
With this technique the longitudinal component of the neutrino momentum $(p^z_\nu)$ can be determined and the energy 
and momentum of the $(K\mu\nu)$ candidate can be calculated. This $(K\mu\nu)$ candidate is then combined with one 
reconstructed track from the primary vertex to determine the $D^{*+}$ mass. The sign of this candidate pion must 
be equal to that of the muon. To  arbitrate between the two possible $(p^z_\nu)$ solutions (two-fold ambiguity) one chooses the 
solution which gives the smallest $D^{*+}$ mass; Monte Carlo (MC) studies show this to be correct $\sim$80\% of 
the time. Finally, the mass difference, $\Delta M = M(D^{*+}) - M(D^0)$, is determined.
The $D^0 \to K^- \mu^+\nu$ analysis begins by considering one kaon candidate (that is a track assumed to be
a kaon) and an opposite sign muon candidate. The muon candidate must satisfy the same cuts used for the {\it trigger muon} of
the $D^0 \to \mu^+\mu^-$ analysis, while the kaon candidate is a track reconstructed in the silicon tracker 
with  $\chi^2/DOF < 3$, $|\eta| < 2.1$ and  $p_T>0.8$ GeV/{\it c}. As in the $D^0 \to \mu^+\mu^-$ analysis, 
the kaon and the muon candidates are required to form a secondary vertex with $\textrm{CL}_\textrm{sec} > 1\%$
and the position of the primary vertex of the event is recomputed excluding the kaon and the muon if they belong to the primary. 
Once the direction between the primary and the secondary vertex is known, the E691 technique described above is used 
to determine the $\nu$ momentum. To form a $D^{*+}$, if more than one pion candidate (that is a track originating
from the primary with $p_T>0.6$ GeV/{\it c}) is found within the $\Delta M$ range ($\Delta M < 180$ MeV/$\textrm{c}^2$), 
that with $\Delta M$ closest to the nominal PDG mass difference is chosen. Due to the charge relation between the $\mu$ and 
the $\pi$ of the decay chain ${D^{*+} \to D^0(K^-\mu^+\nu)\pi^+}$, each $D^0$ candidate selects a {\it Right Sign} (RS) $D^{*}$ 
candidate with $K^-\mu^+\pi^+$ and a {\it Wrong Sign} (WS) $D^{*}$ candidate with $K^-\mu^+\pi^-$.

To reduce prompt background, the separation between primary and secondary vertices was required to be greater
than three times the uncertainty on the separation ($L/\sigma_L > 3$). This cut value was optimized on the 
normalization mode $D^0 \to K^-\mu^+\nu$ to reduce bias.

\begin{figure}[hbtp]
  \begin{center}
    \includegraphics[height=1.5in]{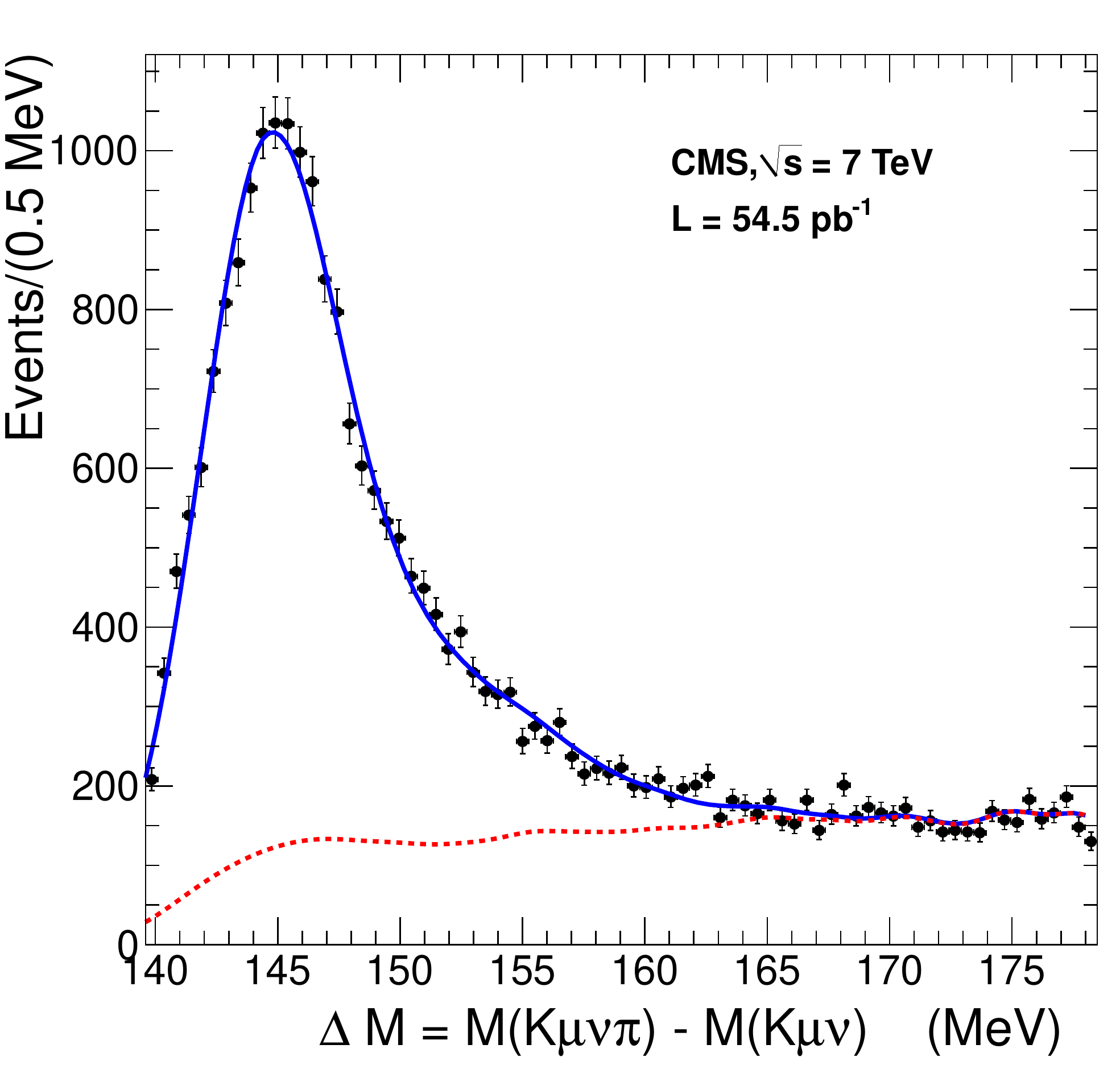}
    \hspace{1cm}
    \caption{$\Delta M = M(K^-\mu^+\nu\pi^+) - M(K^-\mu^+\nu)$ for the RS sample with fit superimposed. 
    The points are the data, the blue line shows the fit function composed of two Gaussians plus the background 
    function modelled by the WS. This background function is shown as a dashed red line.}
    \label{fig:Kmunu}
  \end{center}
\end{figure}

In the WS sample there is no evidence of a signal, as expected. This WS sample is used to model the background 
of the RS sample. Figure \ref{fig:Kmunu} shows $\Delta M = M(K^-\mu^+\nu\pi^+) - M(K^-\mu^+\nu)$ for the 
${D^{*+} \to D^0(K^-\mu^+\nu)\pi^+}$ analysis in the $7^\textrm{th}$ period (the period with the largest dataset). 
The superimposed fit to the unbinned RS data includes two Gaussian functions (the wider one is bifurcated 
to take into account the threshold on $\Delta M$)  with the same mean plus a background function obtained from 
the WS sample. The fit returns $16\,458 \pm 204$ $D^0 \to K^-\mu^+\nu$ candidates.

\begin{figure}[hbtp]
  \begin{center}
    \includegraphics[height=1.5in]{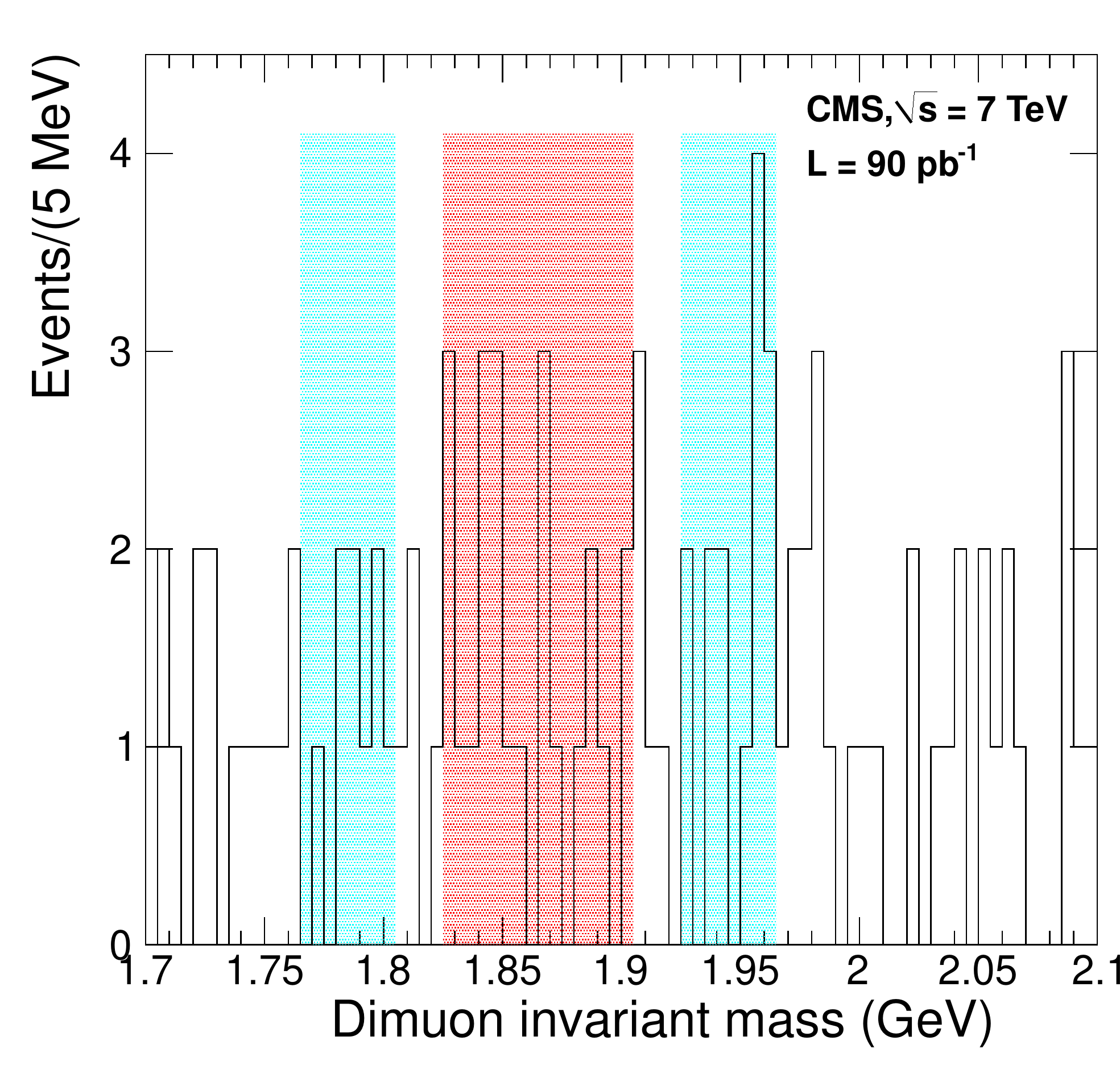}
    \hspace{1cm}
    \caption{The $\mu^+\mu^-$ invariant mass distribution. The light red band shows the signal region, while the two light blue bands show
    the sidebands.}  
    \label{fig:Mumu}
  \end{center}
\end{figure}

\begin{figure}[hbtp]
  \begin{center}
    \includegraphics[width=1.0\textwidth]{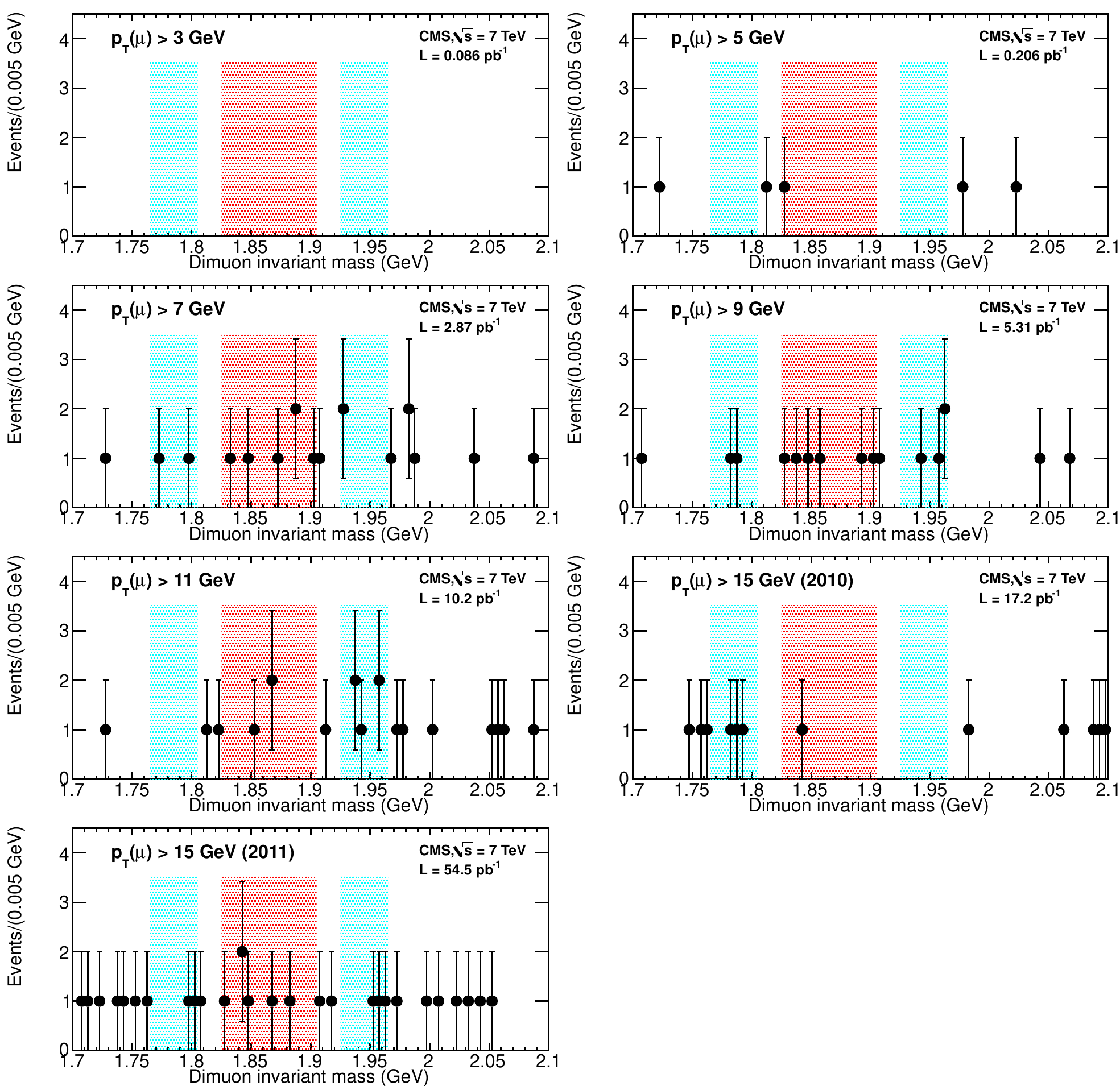}
    \hspace{1cm}
    \caption{The $\mu^+\mu^-$ invariant mass distribution for each trigger period. 
    The light red band shows the signal region, while the two light blue bands show the sidebands.}
    \label{fig:7Mumu}
  \end{center}
\end{figure}

For the $D^0\to\mu^-\mu^+$ analysis , with the additional cut $|\Delta M  - \Delta M_{PDG}| <$ 3 MeV 
where 3 MeV corresponds to $\sim$ 3.6 times the mass resolution measured in MC, we obtain
the $\mu^-\mu^+$ invariant mass shown in Figure \ref{fig:Mumu} (data from all seven periods). 
In more detail, Figure \ref{fig:7Mumu} shows the $\mu^-\mu^+$ invariant mass for each trigger period. 
There is no evidence 
of the $D^0 \to \mu^+\mu^-$ decay.

The upper limit on the branching fraction $D^0 \to \mu^+\mu^-$ is determined using the following formula:

\begin{equation} 
   B(D^0\to \mu^+\mu^-) \leq B(D^0\to K^-\mu^+\nu) \times  \frac{N(\mu\mu)}{N(K\mu\nu)} \times
                        \frac{a(K\mu\nu)}{a(\mu\mu)} \times 
			\frac{\epsilon_{\textrm{trig}}(K\mu\nu)}{\epsilon_{\textrm{trig}}(\mu\mu)}
			\times \frac{\epsilon_{\textrm{rec}}(K\mu\nu)}{\epsilon_{\textrm{rec}}(\mu\mu)}
\label{master_formula}
\end{equation}

where $B(D^0\to K^-\mu^+\nu) = (3.30 \pm 0.13) \times 10^{-2}$ (PDG)~\cite{PDG2010} is the normalization branching fraction,
$N(\mu\mu)$ is the 90\% CL upper limit on the $D^0 \to \mu^+\mu^-$ yield, $N(K\mu\nu)$ is the number of $D^0 \to K^-\mu^+\nu$ candidates, 
and $a$ and $\epsilon$ are the acceptance and the efficiencies of the two modes.

The Monte Carlo simulation is used to determine the acceptance and efficiency ratios for the signal and 
normalization mode. The MC event samples (one for each period) are generated with \textsc{Pythia} 6.409 ~\cite{PYTHIA}, 
the unstable particles are decayed with \textsc{EvtGen} ~\cite{EVTGEN}, and the detector response is simulated 
with \textsc{Geant4} ~\cite{GEANT4}.

The seven periods are simulated with the corresponding triggers and run
conditions, including simulation of the pile-up. The number of MC events in 
each period is proportional to the corresponding statistics of the data.
The average number of reconstructed primary vertices ranges from 1.7 to 5.5 and is well matched by the simulation.
 
The acceptance is defined by $a = N_\textrm{acc}/N_\textrm{gen}$, where $N_\textrm{acc}$ includes events in which
both tracks ($K^-\mu^+$ or $\mu^-\mu^+$) have $|\eta|<2.1$ at the generation level, while the denominator 
$N_\textrm{gen}$ is the total number of signal decays generated. The ratio of acceptance for the two modes
is $a(K\mu\nu)/a(\mu\mu) = 0.995 \pm 0.006$ (stat.).

The trigger efficiency is defined by $\epsilon_\textrm{trig} = N_\textrm{trig}/N_\textrm{acc}$, where $N_\textrm{trig}$ is the 
number of events that pass a particular trigger. Table \ref{table:ratio} shows the ratio of the trigger efficiencies
for $D^0 \to K^-\mu^+\nu$ and $D^0 \to \mu^+\mu^-$ for each of the seven periods. As charm is produced at low $p_T$, 
the trigger efficiencies are very low, especially for the normalization mode.

The reconstruction efficiency is defined by $\epsilon_\textrm{rec} = N_\textrm{rec}/N_\textrm{trig}$, where 
$N_\textrm{rec}$ is the reconstructed signal yield after all the analysis cuts. Table \ref{table:ratio} shows the 
ratio of the reconstruction efficiencies for $D^0 \to K^-\mu^+\nu$ and $D^0 \to \mu^+\mu^-$ in the $7$ periods considered. 
The ratio $\epsilon_\textrm{rec}(K\mu\nu)/\epsilon_\textrm{rec}(\mu\mu)$, ranging from $2.2$ to $0.8$, 
is highly dependent on the trigger.

\begin{table}[htbH]
\begin{center}
\caption{Ratio of trigger and reconstruction efficiencies in the seven periods used in the analysis, uncertainties
are statistical only.}
\label{table:ratio}
\begin{tabular}{|l|c|c|}
\hline
Trigger       & $\epsilon_\textrm{trig}(K\mu\nu)/\epsilon_\textrm{trig}(\mu\mu)$ & $\epsilon_\textrm{rec}(K\mu\nu)/\epsilon_\textrm{rec}(\mu\mu)$ \\
\hline
\hline
HLT\_Mu3      & $0.149 \pm 0.002$    & $2.215 \pm 0.197$ \\
HLT\_Mu5      & $0.112 \pm 0.002$    & $1.651 \pm 0.128$ \\  
HLT\_Mu7      & $0.102 \pm 0.003$    & $1.268 \pm 0.152$ \\  
HLT\_Mu9      & $0.099 \pm 0.003$    & $1.018 \pm 0.097$ \\
HLT\_Mu11     & $0.097 \pm 0.003$    & $0.947 \pm 0.069$ \\  
HLT\_Mu15 (2010) & $0.085 \pm 0.003$ & $0.844 \pm 0.088$ \\  
HLT\_Mu15 (2011) & $0.087 \pm 0.002$ & $0.814 \pm 0.048$ \\ 
\hline
\end{tabular}
\end{center}
\end{table}

\begin{figure}[hbtp]
  \begin{center}
    \includegraphics[width=1.0\textwidth]{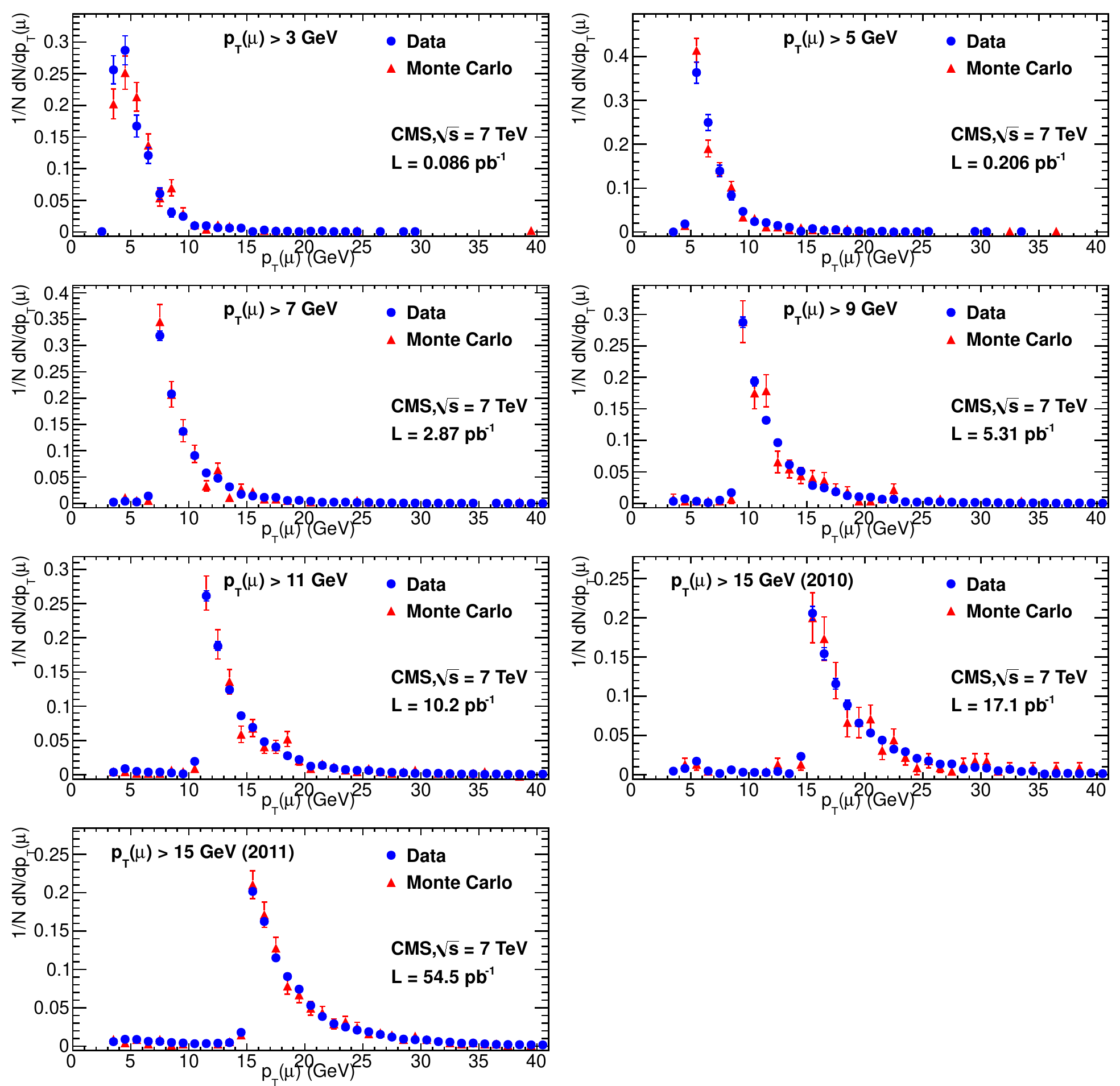}
    \hspace{1cm}
    \caption{Comparison between data and MC of the muon transverse momentum for the $D^0(K^-\mu^+\nu)\pi^+$ analysis in each trigger period.}  
    \label{fig:muPt}
  \end{center}
\end{figure}

To correctly determine the efficiency and acceptance using the MC simulation, the kinematic distributions
in the MC simulation should match the data.
As an example Figure \ref{fig:muPt} shows a comparison between data and MC of the muon $p_T$
for the $D^0(K^-\mu^+\nu)\pi^+$ analysis in each trigger period.

Since there is no clear evidence of $D^0 \to \mu^+\mu^-$ an upper limit on $N(\mu\mu)$ is determined assuming that the number 
of events found in the signal region is the sum of signal and background events (both obeying Poisson statistics).

The hadronic decays $D^0 \to K^-\pi^+$ and $D^0 \to \pi^+\pi^-$ can result in an extra contribution to the low mass sideband 
and signal region, respectively.  This potential contamination was measured in data using reconstructed $D^0\to K^-\pi^+$ decays, 
the measured CMS misidentification rate $\sim 0.01 \%$ \cite{CMS_mupi}, and, for the $D^0 \to \pi^+\pi^-$ decay, the
known relative branching ratio~\cite{PDG2010}. It is found that the hadronic $D^0$ decays produce a negligible contribution 
to the signal and sideband regions.

From Figure \ref{fig:Mumu} it is clear that the background in this limited mass region can be assumed to be flat,
and an estimate can be made from the sidebands.
Three different regions are defined corresponding to the signal region, $M(D^0)_{PDG} \pm 40$ MeV, where $40$ MeV corresponds to 
$\sim 2.1$  times the mass resolution measured in MC, and two sidebands ($40$ MeV wide). The ranges of these three regions are: $1.825 < M_{\mu\mu} < 1.905$ 
for the signal region, $1.765 < M_{\mu\mu} < 1.805$ for the left sideband and $1.925 < M_{\mu\mu} < 1.965$ for 
the right sideband. We find 23 background events ($N_{\textrm{bkg}}$) and 23 events observed 
in the signal region ($N_{\textrm{obs}}$). 
In more detail the $D^0 \to K^- \mu^+\nu$ yields and the number of $\mu^+\mu^-$ events in the signal and sideband
regions are shown for each period in Table \ref{table:yields}.

\begin{table}[htbH]
\begin{center}
\caption{Yields for $D^0 \to K^- \mu^+\nu$, $N_{\textrm{obs}}$ and  $N_{\textrm{bkg}}$ for 
$D^0 \to \mu^+\mu^-$, and systematic uncertainty in the seven periods.}
\label{table:yields}
\begin{tabular}{|l|c|c|c|}
\hline
Trigger       & Y($D^0 \to K^-\mu^+\nu)$ & $N_{\textrm{obs}}$, $N_{\textrm{bkg}}$ & Systematic uncertainty \\
\hline
\hline
HLT\_Mu3      & $ 2412 \pm  145$         & $0, ~ 0$	  & $19.1 \%$ \\
HLT\_Mu5      & $ 2447 \pm  357$         & $1, ~ 0$	  & $18.0 \%$ \\  
HLT\_Mu7      & $11799 \pm  215$         & $6, ~ 4$	  & $19.0 \%$ \\  
HLT\_Mu9      & $ 9982 \pm  176$         & $6, ~ 6$	  & $17.4 \%$ \\
HLT\_Mu11     & $10079 \pm  185$         & $3, ~ 5$	  & $16.4 \%$ \\  
HLT\_Mu15 (2010) & $ 5302 \pm 118$       & $1, ~ 3$       & $17.9 \%$ \\  
HLT\_Mu15 (2011) & $16458 \pm 204$       & $6, ~ 5$       & $15.6 \%$ \\ 
\hline
\end{tabular}
\end{center}
\end{table}

The limit on the $D^0 \to \mu^+\mu^-$ branching fraction depends on statistical and systematic uncertainties.
Several sources of systematic uncertainties have been considered. 
There is no systematic uncertainty on the production cross section, as we use the ratio:
$\frac{D^{*+} \to D^0(\mu^-\mu^+)\pi^+}{D^{*+} \to D^0(K^-\mu^+\nu)\pi^+}$.
The statistical errors on the determination of the acceptance and of the reconstruction efficiency ratio
$\epsilon_\textrm{rec}(K\mu\nu)/\epsilon_\textrm{rec}(\mu\mu)$ (see Table \ref{table:ratio}) are 
taken as a systematic uncertainties. The uncertainties on the reconstruction efficiency of kaons and muons have
been determined using data and found to be $3.9\%$ for the hadron tracking efficiency and $1\%$ for the muon
tracking efficiency \cite{CMS_reckmu}. 
The biggest systematic uncertainty comes from the determination of 
the trigger efficiency, $\epsilon_\textrm{trig}$. The reason is that the $p_T$ spectrum of muons coming from  
charm meson decays is not matched by the CMS triggers. The consequence is an extremely low 
$\epsilon_\textrm{trig}$, especially for the normalization decay mode, which can be a significant 
source of systematic uncertainty. To estimate this contribution we compute the ratio 
$R_\textrm{trig} = (\textrm{Yield}_\textrm{Data}/\textrm{Luminosity}) / \epsilon_\textrm{tot}$ 
for the different triggers. If the Monte Carlo correctly simulates the trigger, the ratio $R_\textrm{trig}$ 
should not depend on the different HLT triggers. The weighted average of the seven periods is calculated as well 
as the PDG scale factor S~\cite{PDG2010}, $S = \sqrt{\chi^2/(\textrm{DOF}-1)} = 1.131$. This S-factor tells how much any 
single error should be increased to have $\chi^2/(\textrm{DOF}-1)=1$. We estimate an uncertainty of 
$13.1\%$ for this source. Variations of the fitting functions used to obtain the yield of the normalization mode gives
systematic uncertainties of $1 \%$ to $9 \%$ for the seven run periods.
Another source of systematic uncertainties is the contamination from other decay modes to the 
Yield($D^{*+} \to D^0(K^-\mu^+\nu)\pi^+$). We consider $D^0 \to K^{*-}(K^-\pi^0)\mu^+\nu$ which is the largest 
contamination for the semileptonic decay $D^0 \to K^-\mu^+\nu$. Using a MC simulation we estimate a contribution of 
$1.8 \%$ from $D^0 \to K^{*-}(K^-\pi^0)\mu^+\nu$. Finally, the uncertainty on the PDG value of 
the branching fraction $B(D^0\to K^-\mu^+\nu)$ is included.
Adding these contributions in quadrature we obtain the systematic uncertainty for each period shown in 
Table \ref{table:yields}, which has been included in the determination of the upper limit.

\section{Conclusions}

The $90 \%$ confidence level upper limit is computed using the $CL_s$ approach ~\cite{Read,Junk}, 
combining the results of the 7 periods. The values used are those reported in Table \ref{table:yields} 
as well as $\epsilon_\textrm{trig}(K\mu\nu)/\epsilon_\textrm{trig}(\mu\mu)$ and 
$\epsilon_\textrm{rec}(K\mu\nu)/\epsilon_\textrm{rec}(\mu\mu)$ as shown in Table \ref{table:ratio}.

The final result is:

\begin{equation} 
   B(D^0\to \mu^+\mu^-) \leq 5.4 \times 10^{-7} (90\% \, \mathrm{CL}).
\label{final}
\end{equation}  

In summary, the FCNC decay $D^0 \to \mu^+\mu^-$ has been searched 
for using the CMS detector. No evidence has been found in $\sim 90 \textrm{pb}^{-1}$ of data.
We show the upper limit at $90 \%$ confidence level in Table \ref{table:experiments}, 
together with the present published best limits. Although this upper limit is not the best limit 
for this FCNC decay, it is the first time a semileptonic decay has been used as the normalization.

\begin{table}[htbH]
\begin{center}
\caption{Upper limit at $90\% \, CL$ for $D^0 \to \mu^+\mu^-$.}
\label{table:experiments}
\begin{tabular}{|l|c|}
\hline
Experiment & Upper limit at $90\% \, CL$ \\
\hline
\hline
BABAR~\cite{BABAR} &  $< 1.3 \times 10^{-6}$ \\
CDF~\cite{CDF1}    &  $< 2.1 \times 10^{-7}$ \\
BELLE~\cite{BELLE} &  $< 1.4 \times 10^{-7}$ \\
this measurement   &  $< 5.4 \times 10^{-7}$ \\
\hline
\hline
\end{tabular}
\end{center}
\end{table}

\end{document}

%% file: econfmacros.tex
\def\beq{\begin{equation}}
\def\eeq#1{\label{#1}\end{equation}}
\def\eeqn{\end{equation}}

\def\beqa{\begin{eqnarray}}
\def\eeqa#1{\label{#1}\end{eqnarray}}
\def\eeqan{\end{eqnarray}}

\let\bar=\overbar

\def\Dslash{\not{\hbox{\kern-4pt $D$}}}
\def\dslash{\not{\hbox{\kern-2pt $\del$}}}

\def\msb{{\bar{\ssstyle M \kern -1pt S}}}